\documentclass[twoside]{article}

\usepackage{PRIMEarxiv}

\usepackage[utf8]{inputenc} 
\usepackage[T1]{fontenc}    
\usepackage{hyperref}       
\usepackage{url}            
\usepackage{booktabs}       
\usepackage{amsfonts}       
\usepackage{nicefrac}       
\usepackage{microtype}      
\usepackage{lipsum}
\usepackage{fancyhdr}       
\usepackage{graphicx}       
\graphicspath{{media/}}     
\usepackage{amsmath} 

\usepackage{enumitem}
\usepackage{soul}
\usepackage{multirow}
\usepackage{multicol}
\usepackage{xcolor}

\newcommand{\wu}[1]{{\color{black}#1}}

\title{Can citations tell us about a paper's reproducibility? A case study of machine learning papers}

\author{
  Rochana R. Obadage \\
  Old Dominion University \\
  Norfolk, VA, USA \\
  \texttt{oruma001@odu.edu} \\
  \And
  Sarah M. Rajtmajer \\
  IST, Pennsylvania State University \\
  University Park, PA, USA \\
  \texttt{smr48@psu.edu} \\
  \And
  Jian Wu \\
  Old Dominion University \\
  Norfolk, VA, USA \\
  \texttt{j1wu@odu.edu} \\
}

\begin{document}
\maketitle

\begin{abstract}
The iterative character of work in machine learning (ML) and artificial intelligence (AI) and reliance on comparisons against benchmark datasets emphasize the importance of reproducibility in that literature. Yet, resource constraints and inadequate documentation can make running replications particularly challenging. 
  Our work explores the potential of using downstream citation contexts as a signal of reproducibility. 
  We introduce a sentiment analysis framework applied to citation contexts from papers involved in Machine Learning Reproducibility Challenges in order to interpret the positive or negative outcomes of reproduction attempts. Our contributions include training classifiers for reproducibility-related contexts and sentiment analysis, and exploring correlations between citation context sentiment and reproducibility scores. 
  Study data, software, and an artifact appendix are publicly available at \textit{\url{https://github.com/lamps-lab/ccair-ai-reproducibility}}.
\end{abstract}


\section{Introduction}
In the rapidly advancing fields of machine learning (ML) and artificial intelligence (AI), emerging technologies are often advancements, refinements, or assemblies of existing ones. Because of this, reproducibility is central to progress in these fields. Yet, the growing complexity of ML research, hardware and software resource constraints, code and data with insufficient documentation, proprietary datasets, and current incentive structures have made the direct reproduction of existing models and findings infeasible in many cases. Indeed, the ML/AI community is beginning to demonstrate that it too faces a crisis of reproducibility \cite{Raff-NEURIPS2019_c429429b,belz-etal-2021-systematic}. Directly reproducing reported results is the most reliable method to test the reproducibility of a paper, but this method does not scale to millions of research papers. Existing efforts to automatically assess reproducibility and replicability either use shallow features directly extracted from paper content, such as bibliographic, statistical, and semantic features \cite{wu2021-SBS}, or latent features such as word embeddings \cite{yang-brian-uzzi}. As yet, no automatic approach to reproducibility or replicability assessment has demonstrated good enough performance to be useful in real-world scenarios.

\begin{figure}[h]
\vspace{-0.1cm}
  \centering
    \setlength{\fboxsep}{0pt} 
    \setlength{\fboxrule}{0.5pt} 
  \fbox{\includegraphics[width=0.6
  \linewidth]{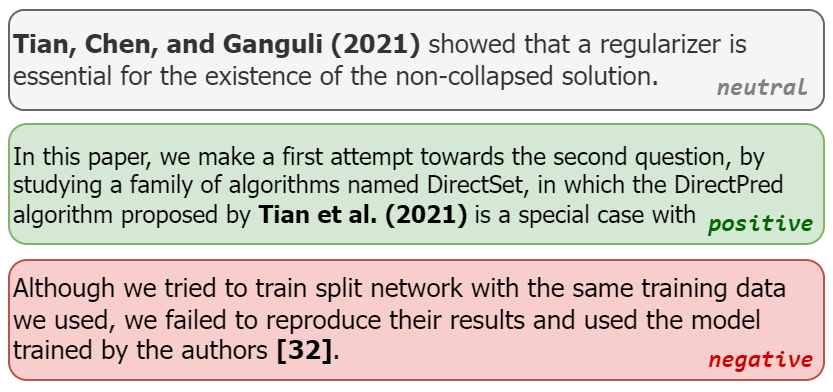}}
  \caption{Examples of citation context with different reproducibility sentiments.}
  \label{citaion_contexts_example}
\vspace{-0.3cm}
\end{figure}

Our work explores opportunities to mine the text around downstream citations of a paper--\textbf{{citation context}} (Figure \ref{citaion_contexts_example})--for cues indicating that an author has successfully or unsuccessfully replicated another paper's work in the course of their own. This is motivated by an understanding that most reproductions and replications are relatively informal (vs. explicitly framed as a replication study) and occur commonly in ML/AI in service to model comparisons against benchmarks, use of one technology in service to a different research aim, or similar. The reproducing or replicating author will note this in their own work, information which--if reliably extracted--could be a rich resource for understanding reproducibility and replicability in the field. 
Although prior research has performed classification on citation contexts for various purposes \cite{Inline-Citation-Classification,SORIANO-MARTÍNEZ-BARCO-2017,meta-semantic-classification,te-etal-2022-citation}, whether citation context can be used as a signal of reproducibility has not yet been studied. In this paper, we address this question by exploring the correlation between reproducibility scores and reproducibility sentiment of citation context. \footnote{We adopt the definition of reproducibility from 
\cite{national2019reproducibility} where a finding is deemed reproducible if consistent results are obtained using the same input data, computational steps, methods, and code, and conditions of analysis. This definition is consistent with the definitions of reproducibility adopted by ACM \cite{acmartifact}.}

Specifically, we train ML models to classify citation contexts based on their reproducibility-based sentiment. This approach falls under the umbrella of aspect-based sentiment analysis, a text analysis technique that classifies text by sentiment related to a given aspect \cite{nazir2022sentiment}. Reproducibility scores are calculated based on the reports of direct reproducibility studies. We then study whether the reproducibility score is correlated with the normalized citation context count of a certain sentiment. Our contributions are as follows:
\vspace{-0.3cm}
 \begin{enumerate}
     \item We observe a correlation between reproducibility scores and citation context sentiment in a pilot dataset containing 41,244 citation contexts extracted from 130 ML papers.
     \item We build the first ground truth dataset of direct reproducibility scores of 22 ML papers and 1937 citation contexts, manually labeled into three reproducibility sentiment categories.
     \item We train and validate two ML models to classify citation context into reproducibility sentiments and obtain F1-scores ranging from 0.70 to 0.86.
 \end{enumerate}
\vspace{-0.1cm}

\section{Related Work}

Our study is related to prior work on reproducibility of ML and AI. While Gundersen et al. \cite{Gundersen_Kjensmo_2018} discuss state-of-the-art of reproducibility in AI, Raff \cite{Raff-NEURIPS2019_c429429b} evaluates the reproducibility of 255 papers published between 1984 to 2017, emphasizing the importance of factors beyond code availability. Akella et al. \cite{Akella} discuss factors that contribute to 
reproducibility of ML findings and propose solutions. 

We build upon work in sentiment analysis. Yousif et al. \cite{Yousif2019MultitaskLM} propose a multitask learning model based on convolutional and recurrent neural networks that perform the citation sentiment and purpose classification. HuggingFace \cite{huggingfaceModelsHugging} contains a repository of open-source ML models for sentiment analysis tasks trained on various datasets. \wu{We used both supervised and unsupervised models trained on different datasets including tweets, social media posts, and citation contexts. }

In terms of citation context classification, Cohan et al. \cite{cohan-etal-2019-structural} propose ``structural scaffolds", a multitask model incorporating structural information of scientific papersfor classification of citation intent. Te et al. \cite{te-etal-2022-citation} compare methods for classification of critical vs. non-critical citation contexts. Budi et al. discuss citation meaning using sentiment, role, and citation function classifications \cite{Budi2023}.

\section{Dataset}\label{Dataset}
\begin{figure}[h]
  \vspace{-0.1cm}
  \centering
    \setlength{\fboxsep}{0pt} 
    \setlength{\fboxrule}{0.5pt} 
  \fbox{\includegraphics[width=0.65\linewidth]{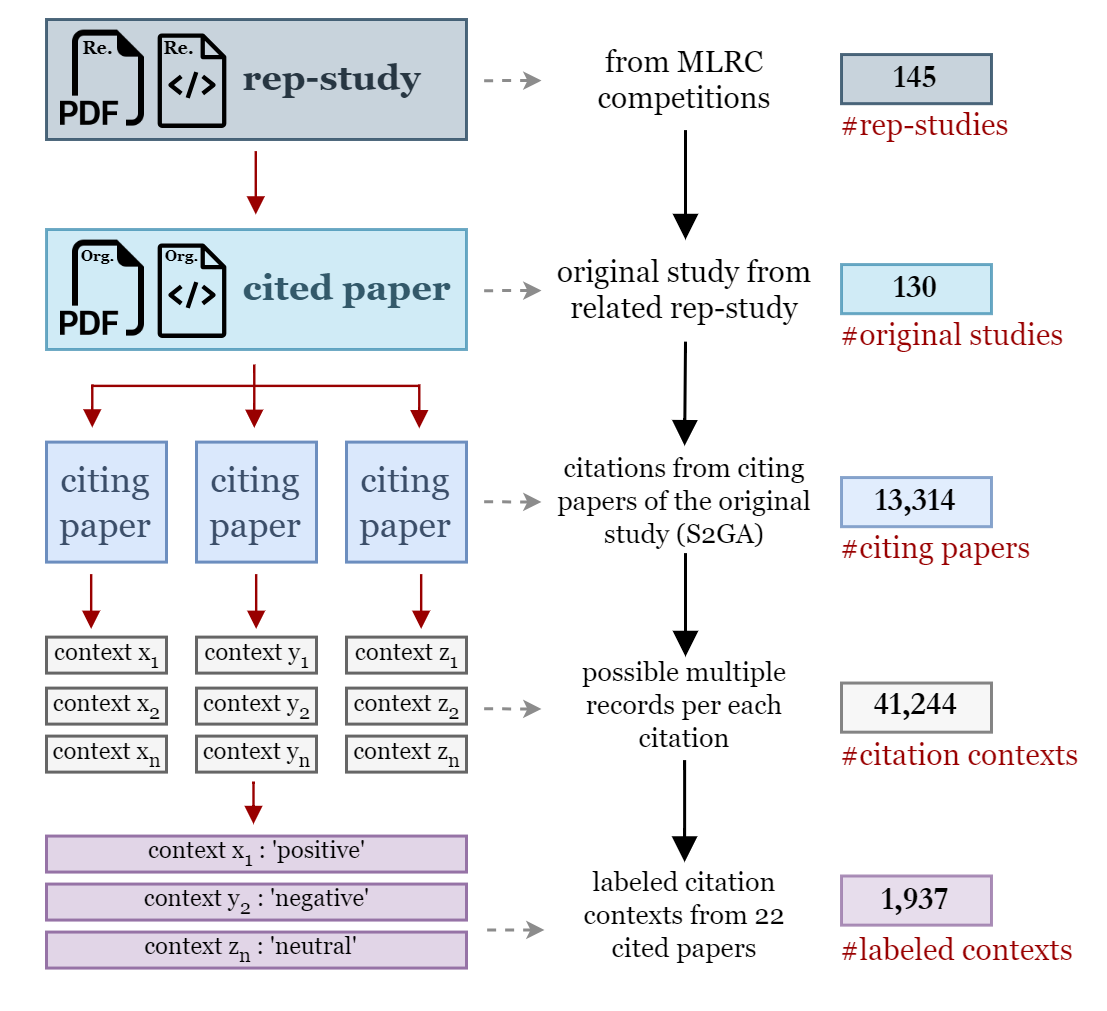}}
  \caption{A schematic illustration of the data reduction and processing workflow. }
  \label{citaion_contexts_relationships}
  \vspace{-0.2cm}
\end{figure}

The dataset contains three types of scientific documents: \textbf{original studies} - original research papers/findings which serve as targets for a reproduction; \textbf{reproducibility studies (rep-studies)} - papers reporting attempts to reproduce a particular paper/finding; and, \textbf{citing papers} - papers that cite the original studies. We prepared our dataset in five steps (see Figure \ref{citaion_contexts_relationships}).
\begin{enumerate}
    \item \wu{Collect reproducibility studies from existing data sources }
    \item Collect metadata for original studies
        \item Calculate the reproducibility score for each rep-study
    \item Collect citation contexts from citing papers
    \item Label citation contexts by reproducibility sentiments 
\end{enumerate}

\subsection{Reproducibility Studies}\label{Reproducibility-Studies}
We collected the metadata for 145 rep-studies from existing data sources listed in Table \ref{tab:rep_study_sources}. The Machine Learning Reproducibility Challenges (MLRC) contains 129 rep-studies of 114 papers (some rep-studies were conducted on the same original paper). Because the majority of papers were successfully reproduced, we supplement the data with 16 rep-studies by Ajayi et al. \cite{10.1007/978-3-031-41679-8_1} including {5} successful and {11} unsuccessful rep-studies. \wu{The same input data, computational steps, methods, and analysis conditions were adopted in all rep-studies except that the experiments were conducted by different teams.} Using the DOIs we collected metadata for all the {130} original studies from the Semantic Scholar Graph API (S2GA; \cite{Kinney2023TheSS}).   

\subsection{Reproducibility Score Calculation}
Traditionally, reproducibility scores are defined as dichotomous values, 
indicating reproducibility of a paper based on its primary finding. 
Here, we introduce an extended reproducibility score (rs\_score) for a paper with \emph{multiple} findings, 
given by the equation below. To calculate it, we perused 
full text of replication studies. The rs\_score distribution is shown in Table~\ref{tab:rsscore}.

\begin{equation*}
  \text{\textbf{rs\_score}} = \frac{\text{\# of successfully reproduced findings}}{\text{\# of total findings selected to reproduce}}
\end{equation*}

\begin{table}[ht]
    \caption{Data sources for selected reproducibility studies with the year of reproduction, the numbers of rep-studies and citation contexts for each data source.}
    \centering
    \begin{tabular}{lccrc}
    \toprule
    \textbf{Data Source}  & \textbf{Year} & \textbf{\#Rep-Studies}& \textbf{\#Contexts} \\
    \midrule
        ICLR \cite{repromlReproducibilityChallenge,rescienceReScience} & 2019 & 4 & 2102 \\
        NeurIPS \cite{repromlReproducibilityChallenge,rescienceReScience} & 2019 & 10 & 9908 \\
        MLRC \cite{repromlReproducibilityChallenge,rescienceReScience} & 2020 & 23 & 10798 \\
        MLRC \cite{repromlReproducibilityChallenge,rescienceReScience} & 2021& 47 & 6958 \\
        MLRC \cite{repromlReproducibilityChallenge,rescienceReScience} & 2022 & 45 & 9869 \\
        TSR \cite{10.1007/978-3-031-41679-8_1} & 2023& 16 & 1609\\
        \toprule
        \textbf{Total} & & 145 & 41244\\
    \bottomrule
    \end{tabular}
    \label{tab:rep_study_sources}
\end{table}

\subsection{Citation Context Collection}\label{Citation-Context-Collection}
We collected all citation contexts available for each of 130 original studies from S2GA \cite{Kinney2023TheSS} (Table ~\ref{tab:rep_study_sources}). 
We collected 13,314 citations and 41,244 citation contexts. On average, each original paper was cited more than 3 times per citing paper.

\begin{table}
\caption{Distribution of original papers and citation contexts over rs\_scores. The rows corresponding to zero original papers (and citation context) are not shown. $N_{\rm pos}$ and $N_{\rm pos}$ for models M6 and M7 are defined in Section~\ref{sec:correlation}.}
\centering
\begin{tabular}{ccccccc}
\toprule
   \multirow{2}{*}{\bf rs\_score}  &\multirow{2}{*}{\bf \#paper} & \multirow{2}{*}{\bf \#Context} & \multicolumn{2}{c}{\bf M6} & \multicolumn{2}{c}{\bf M7} \\
   \cmidrule{4-7}
   &&& \multicolumn{1}{c}{$N_{\rm pos}$} & $N_{\rm neg}$ &\multicolumn{1}{c}{$N_{\rm pos}$}&$N_{\rm neg}$\\
   \midrule
    0.0 & 11 &1288&337&98 & 181& 72\\
    0.2 &1&49&17&6 & 2&4\\
    0.4 &2&152 &46&6 & 32&2\\
    0.5 &12&1967&669&147 & 469&114\\
    0.6 &5&284&90&27& 45&10\\
    0.7 &4&140&33&9 & 26&2\\
    0.8 &3&161&31&9 & 29&6\\
    1.0 &89&37203&14521&2064& 9516&1729\\
    \bottomrule
\end{tabular}
\label{tab:rsscore}
\end{table}

\subsection{Building the Ground Truth}\label{Data-Labelling}
As a pilot study, we randomly selected 22 original papers (Table \ref{tab:rep_study_sources}) and 
1937 citation contexts. We manually labeled these citation contexts into 3 reproducibility sentiments:

\begin{itemize}
    \item \textbf{positive}: the context hints about reproducibility (such as re-usage about the cited paper's data/code or the concept);
    \item \textbf{negative}: the context hints about irreproducibility (such as unavailability of the cited paper's data/code or unsuccessful attempts in reproducing);    
    \item \textbf{neutral}: the context simply mentions (cites) the cited paper without any hints about reproducibility.
\end{itemize}

\noindent The ground truth dataset contains 158 positive (8.1\%), 23 negative (1.2\%), and 1756 neutral (89.7\%) citation contexts. As the distribution is skewed, we down-sampled positive and neutral citation contexts to match the number of negative citation contexts, for a balanced ground truth subset containing 69 labeled citation contexts.

\section{Sentiment analysis}
We first performed aspect-based sentiment analysis by classifying citation context into the three sentiments above. To our knowledge, there are no ready-to-use models for this task, so we trained our own ML models using the balanced ground truth subset (Section \ref{Data-Labelling}). For comparison, we selected five pre-trained open-source multiclass sentiment analysis models from \textit{HuggingFace} \cite{huggingfaceModelsHugging} (Table ~\ref{tab:five-models}) based on the popularity. M1-M5 were trained/fine-tuned on social media posts, tweets, or generic datasets. 

We trained two in-house models using our data. The first (\textbf{M6}) leverages DistilBERT \cite{distilbert} 
fine-tuned using our ground truth data. Compared with BERT \cite{devlin-etal-2019-bert}, DistilBERT is lighter, faster, and achieved the top performance for our previous reproducibility-related study \cite{lamia_url_rep}. Our second model is a hierarchical classifier (\textbf{M7}, combining M7.1 and M7.2) to verify our results obtained from M6. In the first step, we trained a binary classifier (M7.1) which classifies citation contexts 
as related to or not related to reproducibility. We used the full ground truth set (1937 labeled citation contexts) and merged positive and negative labels into a category called \textbf{\textit{related}};  neutral labels were categorized as \textbf{\textit{not related}}. In the second step, we fine-tuned a DistilBERT binary classifier (Table \ref{tab:five-models}: \textit{M7.2}) to classify citation contexts labeled as related as either \textbf{\textit{positive}} or \textbf{\textit{negative}}.




\begin{table}
    \caption{Comparison of mean weighted average precision, recall, and F1-scores for M1-M5.}
    \centering
    \begin{tabular}{lcccc}
        \toprule
            \textbf{Model} & \textbf{Domain} & \textbf{mAP} & \textbf{mAR}& \textbf{mAF$_1$}\\
        \midrule
            M1 \cite{hartmann2021} &social media posts & 0.46 & 0.43 & 0.34 \\
            M2 \cite{Seethal-Hugging}&generic dataset & 0.67 & 0.42 & 0.37 \\
            M3 \cite{finiteautomata-Hugging} &tweets & 0.63 & 0.51 & 0.48 \\
            M4 \cite{Souvikcmsa-Hugging} &generic dataset & 0.54 & 0.42 & 0.35 \\
            M5 \cite{SbcBI-Hugging} &generic dataset & 0.39 & 0.48 & 0.41 \\
        \bottomrule
        \end{tabular}
    \label{tab:five-models}
\end{table}
\begin{table}
    \caption{5-fold cross-validation results for M6 and M7.}
    \centering
    \begin{tabular}{lcccc}

        \toprule
            \textbf{Model} & \textbf{Data} & \textbf{mAP} & \textbf{mAR}& \textbf{mAF$_1$}\\
        \midrule
            M6 (sentiment 3-class) & 93 & 0.81 & 0.71 & 0.70 \\
            \midrule
            M7.1 (related/not related) & 362 & 0.83 & 0.82 & 0.82 \\
            M7.2 (positive/negative) & 46 & 0.92 & 0.83 & 0.86 \\
        \bottomrule
        \end{tabular}
    \label{tab:5-fold-cross-validation}
\end{table}

\section{Results}
\subsection{Sentiment Analysis}
The evaluation results of the five baseline methods are shown in Table~\ref{tab:five-models}. These models are evaluated on the balanced ground truth subset consisting of an equal number of positive, negative, and neutral citation contexts. These baseline models were not trained on our data, which explains why they do not perform well. To evaluate M6 and M7, we performed 5-fold cross-validation (Table~\ref{tab:5-fold-cross-validation}) using the ground truth set supplemented with additional positive and neutral samples. M6 and M7 achieve F1-scores of 0.70 and M7 achieved an F1-score of 
0.86, respectively. We note that we were not able to test M6 and M7 on the identical data as M1-M5 because M6 and M7 incorporate some of this data for training. 
Nevertheless we observe M6 and M7 perform significantly better than baselines and achieve reasonably good performance.

\begin{figure}[htbp]
  \centering
    \setlength{\fboxsep}{0pt} 
    \setlength{\fboxrule}{0.5pt} 
    \includegraphics[width=0.99\linewidth]{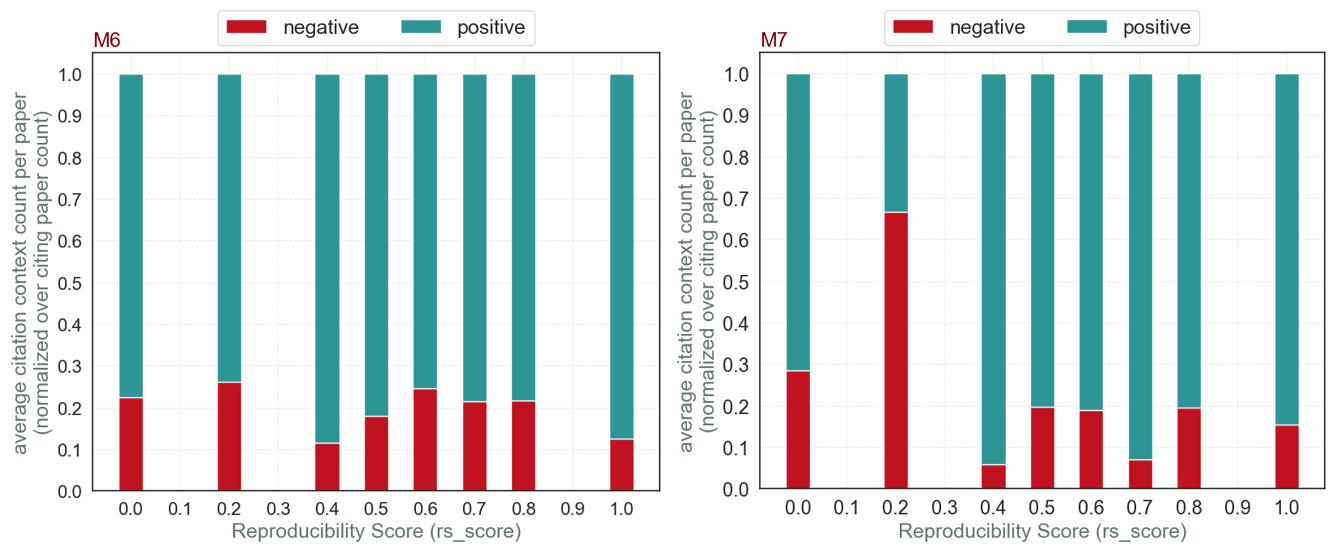}
  \caption{Normalized citation context sentiment counts vs. reproducibility scores using M6 (left) and M7 (right).}
  \label{normalized-Sentiment-Fractions}
\end{figure}

\begin{figure}[htbp]
  \centering
    \setlength{\fboxsep}{0pt} 
    \setlength{\fboxrule}{0.5pt} 
  \includegraphics[width=0.7\linewidth]{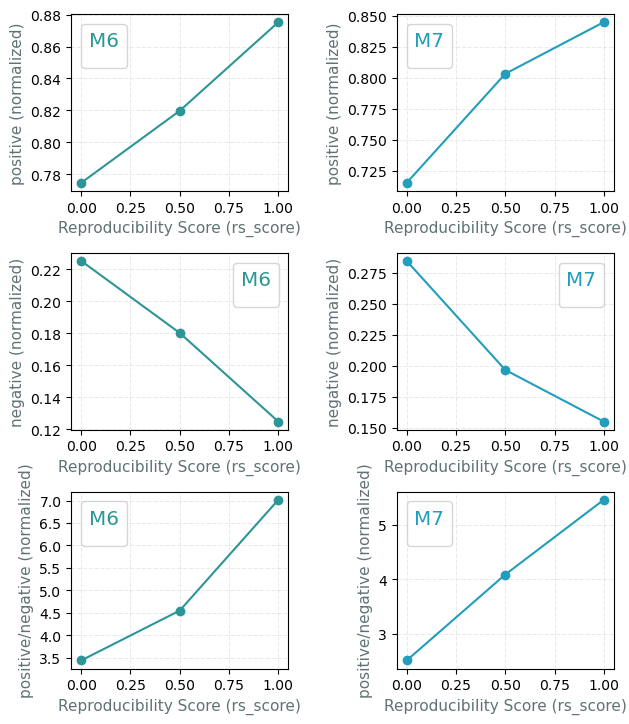}
  \caption{Normalized positive and negative citation context counts 
  vs. rs\_scores using M6 (left) and M7 (right).}
  \label{fig_17}
\end{figure}

Next, M6 and M7 were applied to all 41,244 citation contexts. Applying M6 resulted in 15,744 positive (38.17\%), 2366 negative (5.74\%), and 23,134 neutral (56.09\%) citation contexts. Applying M7 resulted in 10,300 positive (24.97\%), 1939 negative (4.70\%), and 29,005 (70.33\%)  neutral 
citation contexts.

\subsection{Citation Context Sentiments vs. Reproducibility Scores}
\label{sec:correlation}

Our goal is to investigate the correlation between the reproducibility sentiment of citation contexts and the reproducibility scores of original papers. Because of the strongly skewed distribution of reproducibility scores (Table~\ref{tab:rsscore}), simply using the numbers of citation contexts will not result in meaningful conclusions. Therefore, we normalized the number of citation contexts by a factor $Z=N_{\rm pos}+N_{\rm neg}$, 
in which $N_{\rm pos}$ and $N_{\rm neg}$ are the numbers of citation contexts labeled as positive and negative reproducibility-based sentiments, respectively. Therefore, the {\bf normalized citation context counts}, i.e., the fraction of positive or negative citation contexts, are given by: 
$$N^\prime_{\rm pos}=N_{\rm pos}/Z,\quad N^\prime_{\rm neg}=N_{\rm neg}/Z,\quad N^\prime_{\rm pos}+N^\prime_{\rm neg}=1.$$
Figure \ref{normalized-Sentiment-Fractions} depicts $N^\prime_{\rm pos}$ and $N^\prime_{\rm neg}$ using M6 and M7. 

We note that $N^\prime_{\rm pos}$ or $N^\prime_{\rm neg}$ at certain rs\_scores are calculated based on a small number of papers/contexts. For example, there is only one paper whose rs\_score is 0.2 and M7 only predicts 2 positive and 4 negative citation contexts (Table~\ref{tab:rsscore}). The low citation counts in combination with the uncertainty introduced by the sentiment classification models may lead to large uncertainties of $N^\prime_{\rm pos}$ and $N^\prime_{\rm neg}$. To obtain statistically meaningful results, we remove data points calculated based on less than 50 negative citation contexts, leaving three data points at rs\_score = 0, 0.5, and 1. 

Correlations between the normalized citation context count of positive or negative sentiment and the rs\_scores, for M6 and M7, are shown in Figure~\ref{fig_17}. Given a cited paper, the fraction of positive citation contexts ($N^\prime_{\rm pos}$) increases with the reproducibility scores, and the fraction of negative citation contexts ($N^\prime_{\rm neg}$) decreases with the reproducibility scores. The ratio  $r=N^\prime_{\rm pos}/N^\prime_{\rm neg}$ exhibits a magnified correlation with rs\_score, with $r$ ranging from about 3.5 to 7 for M6 and from 2.5 to 5.5 for M7. Because there are only three data points for each diagram, we did not calculate the correlation and regression coefficients. 

\section{Discussion and Conclusion}
In this pilot study, we explored correlations between reproducibility-based sentiments of citation context and reproducibility scores using a total of 41,244 citation contexts. We trained two sentiment analysis models and achieved F1-scores of 0.70--0.86. Both models exhibited an increasing fraction of positive sentiment citation context with rs\_score and a decreasing fraction of negative sentiment citation context with rs\_score. The correlation is stronger in the ratio diagrams. 

\wu{If our findings are verified using larger datasets, it suggests that it is possible to \emph{statistically} estimate the reproducibility of ML papers using downstream citation contexts.} More precisely, our work suggests that 
downstream mentions of a paper contain signals about the efforts ML researchers routinely undertake to reproduce one another's models and findings, often for purposes of extension or comparison, but which are not systematically reported as reproducibility studies. \wu{Nevertheless, this study does not imply to use of citation context sentiments to replace direct experiments to assess reproducibility. 
Rather, they may be useful as a surrogate to study the trends of reproducibility and its correlations with other factors for large corpora of ML papers when direct reproducibility studies are not feasible. }

One limitation is the relatively low number of training data. \wu{In the future, we will extend our labeling to more papers with direct reproducibility studies, such as ML papers with ACM badges and papers with partial reproducibility scores ($0<rs\_score<1$) to verify and confirm our observations. Another potential limitation is the selection bias. Most rep-studies we adopted are from MLRC, which intentionally reproduces papers published in top-tier venues. This bias can be mitigated by collecting more rep-studies based on a homogeneous selection of venues. }




\bibliographystyle{unsrt}  
\bibliography{references}

\end{document}